\documentclass[12pt,english,floatfix,nofootinbib,superscriptaddress,aps,prd,preprint]{revtex4}
\usepackage[utf8]{inputenc}
\usepackage{float}
\usepackage{array}
\usepackage{lipsum}
\usepackage{graphicx}
\usepackage{amsmath,amsthm,amsfonts,amssymb}
\usepackage{graphicx}
\usepackage[english]{babel}
\usepackage{color}
\usepackage{tensor}
\usepackage{esint}
\usepackage[dvips]{epsfig}
\usepackage[dvips]{graphicx}
\usepackage{float}
\usepackage{units}
\usepackage{textcomp}
\usepackage{mathrsfs}
\usepackage{amsmath}
\usepackage[makeroom]{cancel}
\usepackage{amssymb}
\usepackage{amsbsy}
\usepackage{amsfonts}
\usepackage{amssymb,mathrsfs,xcolor}
\usepackage{esint}
\usepackage{array}
\usepackage{graphicx}
\usepackage{mathtools,slashed}
\usepackage{slashed}
\usepackage{natbib}
 \usepackage{booktabs}
 
\usepackage{hyperref}
\hypersetup{
    colorlinks,
    citecolor=blue,
    filecolor=green,
    linkcolor=purple,
    urlcolor=red,
}

\newcommand{\ie}{\begin{equation}}
\newcommand{\fe}{\end{equation}}
\newcommand{\se}{\begin{eqnarray}}
\newcommand{\ff}{\end{eqnarray}}

\usepackage{hyperref}
\hypersetup{colorlinks,breaklinks,
			citecolor=[rgb]{0,0.0,1.0},
            urlcolor=[rgb]{0.0,0.0,1.0},
            linkcolor=[rgb]{0,0.5,0.9}}

\begin{document}

\title{Shadow structure of generalized $k-n$ black-bounce metrics}

\author{Jose R. Nascimento}
\email{jroberto@fisica.ufpb.br}
\affiliation{Departamento de Física, Universidade Federal da Paraíba, Caixa Postal 5008, 58051-970, João Pessoa, Paraíba,  Brazil} 

\author{Ana R. M. Oliveira}
\email{ana.rafaely@academico.ufpb.br}
\affiliation{Departamento de Física, Universidade Federal da Paraíba, Caixa Postal 5008, 58051-970, João Pessoa, Paraíba,  Brazil} 

\author{Albert Yu. Petrov}
\email{petrov@fisica.ufpb.br}
\affiliation{Departamento de Física, Universidade Federal da Paraíba, Caixa Postal 5008, 58051-970, João Pessoa, Paraíba,  Brazil} 

\author{Paulo J. Porfírio}
\email{pporfirio@fisica.ufpb.br}
\affiliation{Departamento de Física, Universidade Federal da Paraíba, Caixa Postal 5008, 58051-970, João Pessoa, Paraíba,  Brazil} 

\author{Amilcar R. Queiroz}
\email{amilcarq@df.ufcg.edu.br}
\affiliation{Departamento de Física, Universidade Federal de Campina Grande, Caixa Postal 10071, 58429-900, Campina Grande, Paraíba, Brazil}


\date{\today}

\begin{abstract}
The  existence of black hole shadows is one of the most interesting effects of the strong field regime of general relativity (GR). Recent observations by the Event Horizon Telescope (EHT) have provided high-resolution images of the vicinity of supermassive black holes, ushering in a new era for testing gravitation on astrophysical scales. In this work, we continue the investigation initiated by \cite{furtado2025gravitational}, focusing on shadows associated with generalized $k-n$ \emph{black-bounce} type spacetimes, which smoothly interpolate between regular black holes and wormholes. We consider a generalization of the metric with free parameters $(a,k,n)$ that modify the mass function and enrich the possible phenomenology. We develop a semi-analytical study of photon orbits, obtaining the critical impact parameter and the shadow radius for different parameter combinations. Subsequently, we perform numerical ray-tracing simulations using the \textsc{GYOTO} code, incorporating optically thick accretion disks and varying the observation angle. Our results reveal characteristic signatures, including the formation of double-ring structures and deformations of the shadow radius, which can serve as observational discriminators between classical black holes and \emph{black-bounce} solutions.
\end{abstract}

\maketitle

\section{Introduction}

The study of  physical phenomena typical for strong gravity regimes has received increasing attention in recent decades, especially following advances in direct observations of compact objects, where GR is tested in extreme conditions \cite{Wald1984,Chetouani1984,Carroll2004}. Among these phenomena, the \emph{black hole shadow} \cite{Chandrasekhar1983,bardeen1972,Shakura1973,Novikov1973} stands out as one of the most remarkable predictions of Einstein's theory, representing the dark silhouette delimited by the photon capture region \cite{Einstein1936,Liebes1964,Refsdal1964,Soares2025,Kumar2023,Heidari2025,Filho2024a}.
The first theoretical discussion on black hole shadows dates back to Synge \cite{Synge1966} in 1966 and was further developed by Bardeen in 1973 in the context of the Kerr metric \cite{Bardeen1973,Kerr1963,Penrose1965}. More recently, in the context of modified theories of gravity, the shadows of new solutions of compact objects (black holes and wormholes) have been discussed  \cite{Yuennan:2025uvc, Shodikulov:2025xax, Mushtaq:2025ekt, Ahmed:2025grn, Naskar:2025cio, Liu:2025brb, Yasir:2025npe}
These pioneering works showed that the shadow contour is defined by the unstable circular photon orbits, i.e., the photon sphere.

Subsequent studies, such as those by Cunningham and Bardeen \cite{Cunningham1973} and DeWitt-Morette \cite{DeWitt1974}, explored the optical appearance of stars and sources around black holes, paving the way for more systematic investigations into gravitational lensing and emission configurations \cite{Refsdal1964,Filho2024a,Filho2024b,Filho2024c,Eiroa2002,Eiroa2004,Liebes1964,Soares2025,Kumar2023}. A decisive milestone occurred in 2019 when the Event Horizon Telescope (EHT) collaboration released the first direct image of the shadow of the supermassive black hole M87* \cite{Akiyama2019,Akiyama2019b,Akiyama2019c,Akiyama2019d,Akiyama2019e,Akiyama2019f,Akiyama2019a,akiyama2019first}, providing unprecedented observational confirmation of GR's predictions in the strong-field regime. More recently, the EHT also observed the shadow of Sagittarius A* \cite{EHT2022}, consolidating this new field of investigation. A panoramic overview of analytical and numerical techniques can be found in \cite{Cunha2018,Perlick2022,Tsukamoto2021,Vagnozzi2022}.

Black hole shadows have been analyzed in various theoretical and numerical contexts, including exact solutions and analytical approximations \cite{Perlick2022}.
Beyond reflecting fundamental properties, such as the mass and spin of a compact object, shadows can carry signatures of extensions of GR, making them valuable tools for testing alternative theories of gravity \cite{Bambi2019,Cardoso2019,Berti2015}.

In this broader context, growing interest has turned to \emph{black-bounce type spacetimes}, a class of regular solutions that interpolate between black holes and wormholes. Initially proposed by Simpson and Visser \cite{SimpsonVisser2019,Fan2016,Nascimento2020,Lobo2021,Tsukamoto2021} and later extended in different formulations, these models replace the central singularity with a regular throat, characterized by the bounce parameter $a$. Depending on the values of the free parameters, the solution can describe a regular black hole, a one-way wormhole, or even a traversable wormhole \cite{Visser1995,Hawking1973,Nandi2006,Dey2008,Bhattacharya2010,Nakajima2012,Tsukamoto2012}.

A notable generalization of these models was presented in \cite{furtado2025gravitational}, where additional parameters $(k,n)$ are introduced into the mass function, expanding the possible phenomenology. This formalism allows for the existence of multiple horizons, horizonless regimes, and transitions to ultra-regular compact objects, connecting black holes, wormholes, and nonsingular geometries within a unified framework. Recent works \cite{Tsukamoto2021,Lu2020} have shown that such spacetimes can produce distinct observational signatures, such as double-ring structures and modified critical radii, which can be tested by high angular resolution observations.

Beyond semi-analytical analyses, numerical simulations using ray-tracing codes, such as \textsc{GYOTO} \cite{Vincent2011,Aimar2024,Misner1973,Chandrasekhar1983}, have become essential tools for generating realistic images of accretion disks in relativistic regimes. The combination of such codes with astrophysical thin-disk models, like that of Page and Thorne \cite{PageThorne1974}, allows connecting theoretical predictions to direct observations made by collaborations like the EHT.

In this work, we present a systematic analysis of shadows in generalized black-bounce spacetimes, initially considered in Ref. \cite{furtado2025gravitational}. We investigate the influence of the bounce parameter $a$ and the exponents $(k,n)$ on the photon sphere radius, the critical impact parameter, and the shadow morphology. We complement the analysis with ray-tracing simulations using the \textsc{GYOTO} code, incorporating realistic emission models from accretion disks and varying the observation angle. Our results reveal a rich phenomenology, including the emergence of double-ring structures and deformations in the shadow radius, offering new perspectives for distinguishing between classical black holes and black-bounce geometries in future astrophysical observations.

The structure of the paper looks like follows. In the Section 2, we present a general discussions of black-bounce metrics and their properties. In the Section 3, we discuss the dependence of the shadow radius on the $a$  parameter of our metric. The Section 4 is devoted to numerical studies of the radiation intensity. In the Section 5, we preform a detailed investigation of shadows with use of GYOTO. Our results are summarized in the section 6.

\section{Shadows of black-bounce metrics}

In Ref. \cite{furtado2025gravitational}, the authors investigated gravitational lensing in black-bounce spacetimes, in particular, they  considered the shadows of these metrics. However, a more detailed discussion of such shadows can be performed, this is the objective of this work. Let us first discuss the general properties of these metrics, which are given by the following line element:

\begin{equation}\label{svmetric}
    ds^2 = f(r)dt^2 - \frac{dr^2}{f(r)} - \Sigma(r)^2\left(d\theta^2 + \sin^2\theta \, d\phi^2\right),
\end{equation}

\begin{equation}
    f(r) = 1 - \frac{2M(r)}{\Sigma(r)},
\end{equation}

\begin{equation}
    \Sigma(r) = \sqrt{r^2 + a^2}.
\end{equation}

For the constant $M(r)$ case, this metric reduces to the well-known Simpson-Visser one \cite{SimpsonVisser2019}. 
 The fundamental characteristic of black-bounce type spacetimes is their dependence on the parameter $a$, so that, depending on its value, a singularity may appear in the system. This parameter introduces a throat at the origin, which may or may not be associated with an event horizon. The nature of this throat — spacelike, lightlike, or timelike — determines the physical properties of the solution, as illustrated by the metric \eqref{svmetric} \cite{lobo2021novel}.

For certain values of $a$, the metric describes a regular black hole, whose spacelike throat connects the current universe to a future copy. If the throat is lightlike, the solution corresponds to a one-way wormhole. If the throat is timelike, a traversable wormhole is obtained.

Although these spacetimes share properties with Simpson-Visser (SV) models, critical differences stand out, such as the possible existence of multiple horizons and the selective violation of GR energy conditions. Additionally, in modified theories of gravitation, such metrics may, in some cases, satisfy all energy conditions, suggesting a distinct physical scenario.

These characteristics make black-bounce spacetimes a promising field for investigating the interface between regular black holes, wormholes, and extensions of classical gravitation.

One of the major differences between the SV metric described by \eqref{svmetric} with constant $M$, and the generalized one is the generalization of the mass function $M(r)$, which carries additional parameters. The example spacetime proposed in \cite{furtado2025gravitational} introduces the parameters $k$ and $n$, defining the mass as

\begin{equation}\label{massa}
M(r) = \frac{m\Sigma(r)r^k}{\left(r^{2n} + a^{2n}\right)^{\frac{k+1}{2n}}}.
\end{equation}

As verified by \cite{furtado2025gravitational}, for the SV case, the metric imposes that $n=1$ and $k=0$, and when $a \xrightarrow{}0$, we return to the spacelike Schwarzschild solution. In this context, light deflection and gravitational lensing effects using the photon sphere analysis with values of $k=0$, $n=2$ and $k=2$, $n=1$ were studied in  \cite{furtado2025gravitational}.

In the weak-field regime, the dynamics of photons in black-bounce type metrics, with a mass function given by \eqref{massa}, can be conveniently derived from the Lagrangian associated with the motion of a massless particle:

\begin{equation}
\mathcal{L} = \left( 1 - \frac{2mr^k}{(r^{2n} + a^{2n})^{{\frac{k+1}{2n}}}} \right) \dot{t}^2 - \left( 1 - \frac{2mr^k}{(r^{2n} + a^{2n})^{{\frac{k+1}{2n}}}} \right)^{-1} \dot{r}^2 - (r^2 + a^{2})(\dot{\theta}^2 + \sin^2\theta \, \dot{\phi}^2). 
\end{equation}

Imposing the condition for null trajectories, $\mathcal{L} = 0$, and exploiting the temporal and axial symmetries of the metric (since it is static and spherically symmetric), we obtain the associated constants of motion: the photon energy, given by $E = \left( 1 - 2M(r) \right) \dot{t}$, and the angular momentum in the azimuthal direction, $L = (r^2 + a^2)\sin^2\theta\, \dot{\phi}$. Restricting the analysis to the equatorial plane, so that one fixes $\theta = \pi/2$, we obtain an effective equation for the radial motion of the form $\dot{r}^2 + V_{\text{eff}}(r) = E^2$, where $V_{\text{eff}}(r)$ represents the effective potential, given by:

\begin{equation}\label{eq:potencial}
V_{\text{eff}}(r) = \left(1 - \frac{2mr^k}{(r^{2n} + a^{2n})^{{\frac{k+1}{2n}}}}\right) \frac{L^2}{(r^2 + a^2)}.
\end{equation}

The study of the properties of this effective potential allows for the identification of the unstable circular photon orbits, which define the so-called "photon ring" \cite{furtado2025gravitational}. These orbits delimit the region of light capture by the compact object and are closely related to the formation of the "black hole shadows" observed by instruments such as the EHT (Event Horizon Telescope) \cite{Misner1973,EHT2019,LIGO2016}.

\section{shadow radius analysis}

In this section, we discuss black hole shadows in the strong-field regime, with an emphasis on the critical radius $r_m$. This region is particularly important as it harbors unstable circular photon orbits, which define the critical light trajectories around the compact object. The shadow itself corresponds to the area where these photons are captured, creating the characteristic dark disk which can be observed.

In the earlier paper by some of us \cite{furtado2025gravitational}, the shadows were briefly analyzed. We here discuss that static metric more comprehensively. For an observer coming from infinity, the shadow radius is given by the equation:

\begin{equation}
    r_s= \frac{r_m}{\sqrt{f(r_m)}}
\end{equation}

For the analyzed metric \eqref{svmetric}, the critical radius is equal to

\begin{equation}
    r_m = 3 \sqrt{m^2 - \left(\frac{a}{3}\right)^2}
\end{equation}

Therefore, the shadow radius can be written as

\begin{equation}\label{eq:rs}
    r_s = 3\sqrt{3} \sqrt{m^2-\left(\frac{a}{3}\right)^2},
\end{equation}

As we have mentioned, black holes correspond to the values of $a$ within the interval $(0 < a < 2m)$ for the Schwarzschild case, while its further increasing yields wormholes. We analyze the parameters of the metric given by equation \eqref{svmetric} for different values of $a$, $k$, and $n$.

\begin{figure}[h!]
    \centering
    \includegraphics[width=0.5\linewidth]{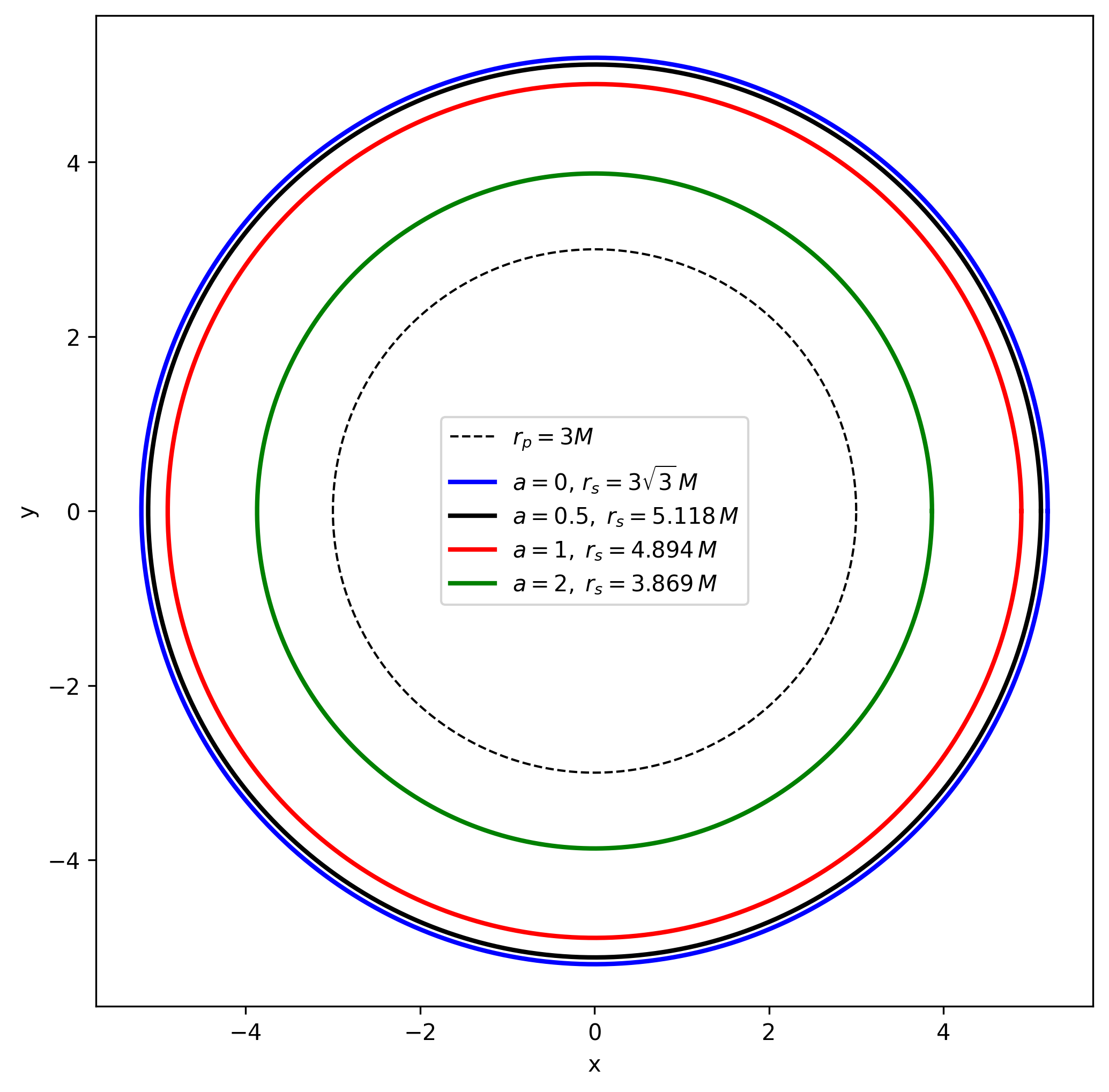}
    \caption{In the figure, we observe the difference in the shadow radius according to different values of $a$.}
    \label{fig:1}
\end{figure}

Figure \ref{fig:1} displays the behavior of the radius within the generalized black-bounce metric, based on Eq. \eqref{fig:1}. In this case, it is observed that as $a$ increases, the radius $(r_s)$ decreases – with an upper limit for the bounce parameter given by $a = 3m$.

We observe that the parameter $a$ lies within the acceptable region for shadows, modifying only the value of the radius. Furthermore, as expected, the largest radius is represented by the Schwarzschild metric, as argued in \cite{lu2020schwarzschild} with $a = 0$.

\section{Shadow Analysis and Intensity Profiles}

In this section, we conduct a detailed analysis of the shadow cast by the black-bounce spacetime, aiming to identify observational signatures that depend on the bounce parameter $a$. The shadow of a black hole is defined by the unstable photon orbit, known as the photon sphere. The shadow radius, $b_{crit}$, which corresponds to the critical impact parameter for an observer at infinity, is related to the photon sphere radius, $r_{ph}$, through the expression:
\begin{equation}
    b_{crit}^2 = \frac{r_{ph}^2}{f(r_{ph})} = \frac{1}{V_{eff}(r_{ph})},
\end{equation}
where $V_{eff}(r) = f(r)/r^2$ is the effective potential for massless particles. The radius $r_{ph}$ is obtained numerically by finding the maximum of this potential.

\begin{figure*}[h!]
    \centering
    \includegraphics[width=\textwidth]{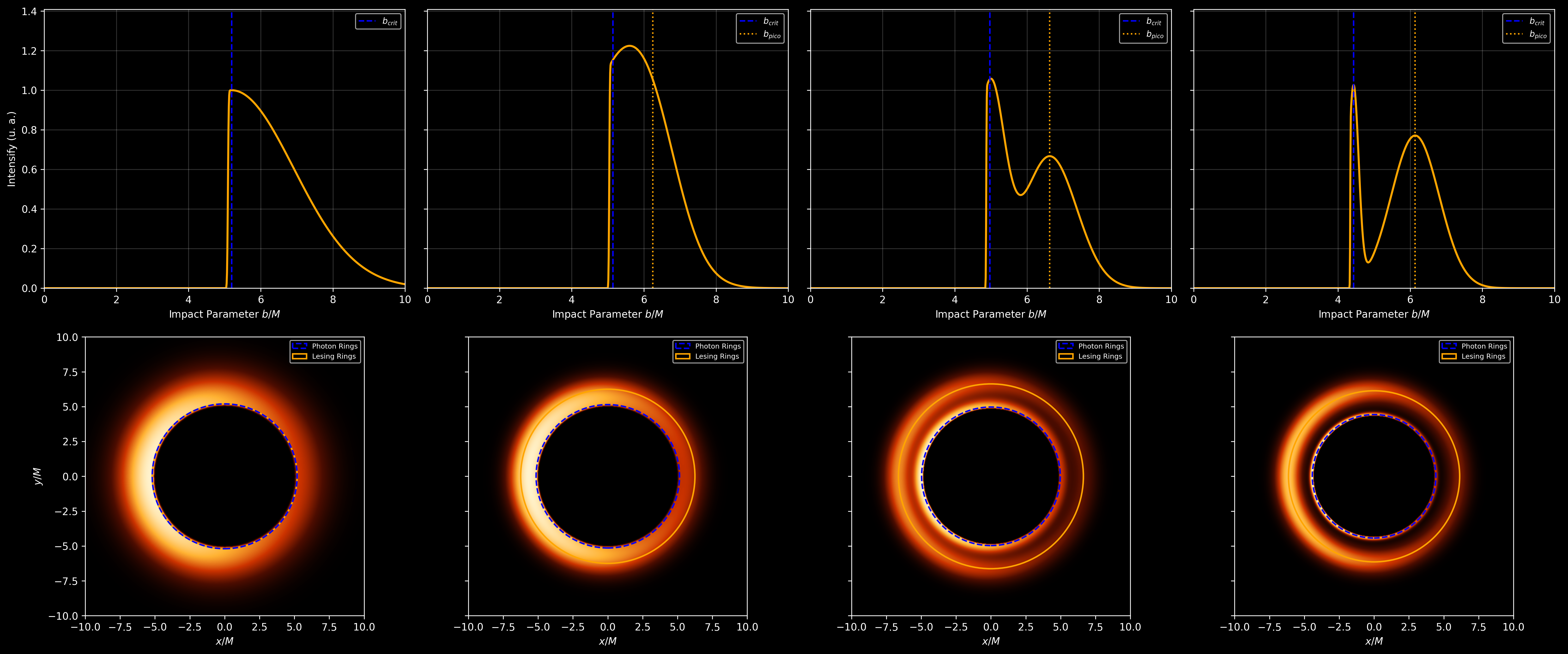}
    \caption{Comparison of the shadow and intensity profile for different values of the bounce parameter $a$, with $k=n=1$ and $m=1$. The top panels show the radial intensity profile, while the bottom panels display the corresponding shadow. (a) The Schwarzschild case ($a=0$). (b) A black-bounce with $a=0.3 \ m$. (c)  A black-bounce with $a=0.6 \ m$. (d) A black-bounce with  $a=1.0 \ m$. The blue dashed lines indicate the photon ring ($b_{crit}$) and the orange solid lines indicate the peak of the lensing ring.}
    \label{fig:comparison_shadows}
\end{figure*}

To visualize the physical implications, we developed a numerical model that simulates the radiation intensity emitted by an optically thin, rotating accretion disk around the compact object. The intensity model is unified, it depends continuously on the parameter $a$, allowing for a smooth transition between the phenomenology of a Schwarzschild black hole and that of a generalized $k-n$ black-bounce. Figure \ref{fig:comparison_shadows} presents the results for the cases where $k=n=1$, varying the parameter $a$ between $0.0 \ m$ and $1.0 \ m$.

The analysis of Figure \ref{fig:comparison_shadows} reveals a rich phenomenology dependent on $a$. Panel (a) corresponds to the Schwarzschild limit ($a=0$). The intensity profile exhibits a single sharp peak, and in the shadow, the photon ring and the lensing ring are degenerate, forming a single bright light ring at the edge of the shadow, whose radius is $b_{crit} = 5.1962 \ m$. This result serves as our reference case.

In panel (b), for a small value of $a=0.3 \ m$, the geometry begins to deviate from the classical case. Although the shadow image is still dominated by a main ring, the intensity profile already reveals the emergence of an incipient second peak, and the shadow radius slightly decreases to $b_{crit} = 5.1502 \ m$.

The transition becomes prominent in panels (c) and (d). For $a=0.6 \ m$, the intensity profile shows two well-defined peaks, which correspond to two visually separate ring structures in the shadow image. The inner ring (blue) remains the photon ring at $b_{crit} = 4.9654 \ m$, while a second, outer lensing ring (orange) becomes visible, with an intensity peak at $b_{peak} = 6.7766 \ m$. For $a=1.0 \ m$, this separation is even more pronounced, with $b_{crit}$ decreasing to $4.8814 \ m$ and $b_{peak}$ moving to $6.8041 \ m$.

The precise numerical results for each case are summarized in Table \ref{tab:resultados}. The quantitative analysis confirms two main trends: (i) the shadow radius, $b_{crit}$, is a monotonically decreasing function of $a$, indicating that the presence of the bounce parameter reduces the size of the dark region; (ii) for $a > 0$, a double-ring structure emerges, whose separation, $\Delta b = b_{peak} - b_{crit}$, increases with $a$.

\begin{table}[h!]
    \centering
    \caption{Numerical values of the shadow radius ($b_{crit}$) and the lensing ring peak position ($b_{peak}$) for the analyzed cases, with $m=1$, $n=1$, and $k=1$, in units of $m$.}
    \begin{tabular}{c c c}
        \hline
        \hline
        
Parameter $a$ &Shadow Radius $b_{crit}$ & Lensing Ring Peak $b_{peak}$\\
        \hline
        0.0 & 5.1962 & 5.1962 \\
        0.3 & 5.1502 & 6.2238 \\
        0.6 & 4.9723  & 6.6303 \\
        1.0 & 4.8814 & 6.8041 \\
        \hline
        \hline
    \end{tabular}
    \label{tab:resultados}
\end{table}

This double-ring structure constitutes a clear observational signature, potentially capable of distinguishing a generalized $k-n$ black-bounce spacetime from a Schwarzschild black hole through high angular resolution observations, such as those conducted by the Event Horizon Telescope.

\subsection{Influence of the parameter $k$}

\begin{figure*}[h!]
    \centering
    \includegraphics[width=\textwidth]{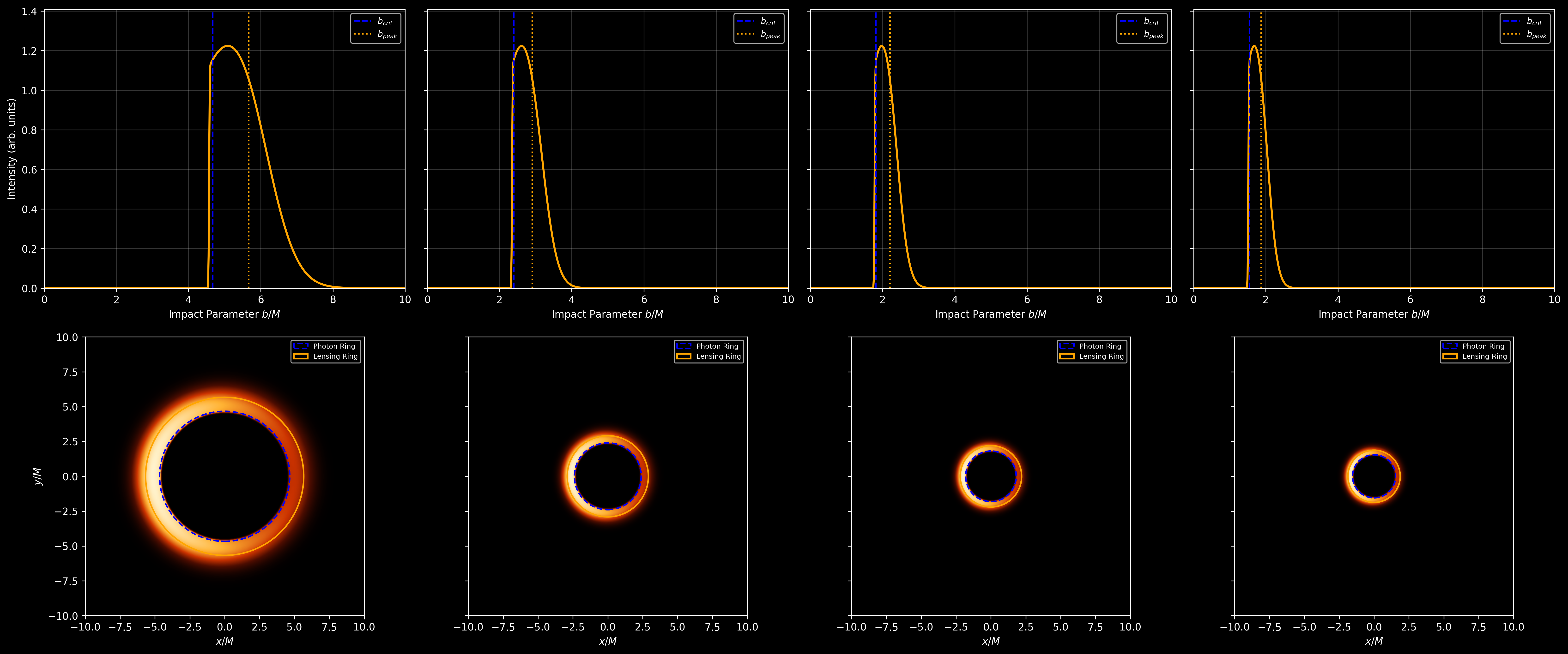} 
    \caption{Comparison of the shadow and intensity profile for different values of parameter $k$, with $a=0.3 \ m$ and $n=1$ fixed. Increasing the value of $k$ results in a progressive decrease of the shadow radius, $b_{crit}$, and, consequently, of the lensing ring radius, $b_{peak}$.}
    \label{fig:comparison_k}
\end{figure*}

To isolate the impact of the exponent $k$ on the shadow geometry, we conducted a second numerical analysis. In this stage, we kept the mass parameter $m=1$, the bounce parameter $a=0.3 \ m$, and the exponent $n=1$ fixed, while $k$ varies between 1.0 and 4.0. The parameter $k$, according to the metric, modulates how the mass $M$ influences the curvature, acting as an exponent on the modified radius term $(r^n + a^n)$. The results of this analysis are compiled in Figure \ref{fig:comparison_k} and Table \ref{tab:resultados_k}.

Figure \ref{fig:comparison_k} demonstrates a clear dependence of the shadow size on the value of $k$. We observe that as $k$ increases, the shadow radius, $b_{crit}$, consistently decreases. For $k=1.0$, we obtain the largest shadow radius in this set, $b_{crit} = 4.9654 \ m$. When increasing $k$ to $4.0$, the shadow radius is reduced to $4.6865 \ m$. This behavior indicates that higher values of $k$ effectively "weaken" the gravitational field in the photon sphere region, allowing light to pass closer to the central object without being captured.

It is important to note that, since our visual model for the lensing ring depends on the position of the photon ring ($b_{peak} \propto b_{crit}$), the radius of the lensing ring also decreases as $k$ increases. However, the qualitative double-ring structure, which is characteristic of the black-bounce spacetime and controlled by the fixed parameter $a$, is preserved in all panels. The precise numerical values are presented in Table \ref{tab:resultados_k}.

\begin{table}[h!]
    \centering
    \caption{Numerical values of the shadow radius ($b_{crit}$) and the position of the lensing ring peak ($b_{peak}$) as a function of parameter $k$, for $m=1$, $a=0.3 \ m$, and $n=1$, in units of $m$.}
    \begin{tabular}{c c c}
        \hline
        \hline
        Parameter $k$ & Shadow Radius $b_{crit}$ & Lensing Ring Peak $b_{peak}$ \\
        \hline
        1.0 & 4.9654 & 6.7766 \\
        2.0 & 4.8020 & 6.5532 \\
        3.0 & 4.7262 & 6.4497 \\
        4.0 & 4.6865 & 6.3955 \\
        \hline
        \hline
    \end{tabular}
    \label{tab:resultados_k}
\end{table}

In summary, while the parameter $a$ governs the existence and separation of the double-ring structure, the parameter $k$ primarily acts as a modulator of the overall shadow size.

\subsection{Parameter Influence and Limits of $n$}

Finally, we investigated the influence of the exponent $n$ on the shadow structure. For this analysis, we fixed the parameters $m=1$, $a=0.2 \ m$, and $k=2.0$, and $n$ varies in the range from $n=1.0$ to $n=4.0$. During this investigation, we encountered a significant physical result: for the fixed parameters, the numerical simulation was unable to compute a shadow for $k=1, 2$, $3$, and $4$.

This negative result is, in fact, an important discovery about the physics of the metric. The impossibility of calculating $b_{crit}$ indicates that, for these parameter combinations, there is no stable and observable photon sphere outside the event horizon. The mathematical condition for the existence of a shadow, $f(r_{ph}) > 0$, is not satisfied due to the limit $k > 2n-1$.

\begin{figure}[h!]
    \centering
    \includegraphics[width=1.0\textwidth]{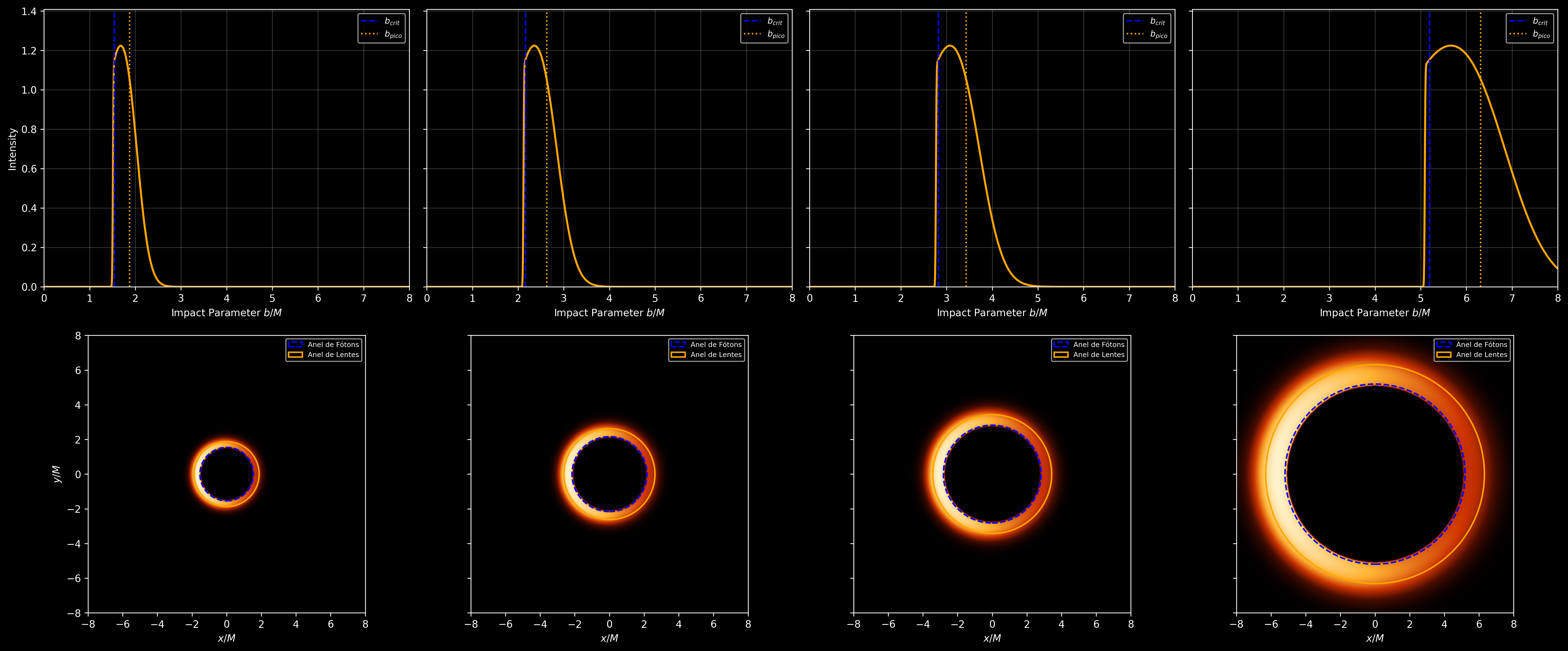} 
    \caption{Simulation of the shadow and intensity profile for the case where an observable shadow is formed, with $n=4.0$ (and $a=0.3 \ m$, $k=4.0$ fixed). The image displays the characteristic double-ring structure of the black-bounce.}
    \label{fig:n_case}
\end{figure}

For $k \geq 4.0$, the shadow reappears, exhibiting the expected characteristics of a black-bounce with $a>0$. The intensity profile shows two peaks, and the shadow is composed of two light rings. Since the value of $a=0.3 \ m$ is small, the separation between the photon ring ($b_{crit}$) and the lensing ring ($b_{peak}$) is minimal, as confirmed by the values in Table \ref{tab:resultados_n}.

\begin{table}[h!]
    \centering
    \caption{Numerical values of the shadow radius ($b_{crit}$) and the position of the lensing ring peak ($b_{peak}$ in units of $m$) as a function of parameter $n$, for $m=1$, $a=0.3 \ m$, and $k=4.0$.}
    \begin{tabular}{c c c}
        \hline
        \hline
        Parameter $n$ & Shadow Radius $b_{crit}$ & Lensing Ring Peak $b_{peak}$ \\
        \hline
        1.0 & 1.5426 & 1.8739 \\
        2.0 & 2.1611 & 2.6254 \\
        3.0 & 2.8220 & 3.4282 \\
        4.0 & 5.1956 & 6.3118 \\
        \hline
        \hline
    \end{tabular}
    \label{tab:resultados_n}
\end{table}

The analysis of the parameter $n$ reveals that $n$ and $k$ are intrinsically related, highlighting their complex and non-linear nature. Unlike $a$, which controls the separation of the rings, and $k$, which modulates the overall shadow size, $n$ appears to govern the very condition for the existence of stable circular light orbits. This result imposes important constraints on the physically viable parameter space for shadow formation in generalized black-bounce spacetimes.


\subsection{Analysis of free parameters}

\begin{figure*}[h!]
	\centering
	\includegraphics[width=\textwidth]{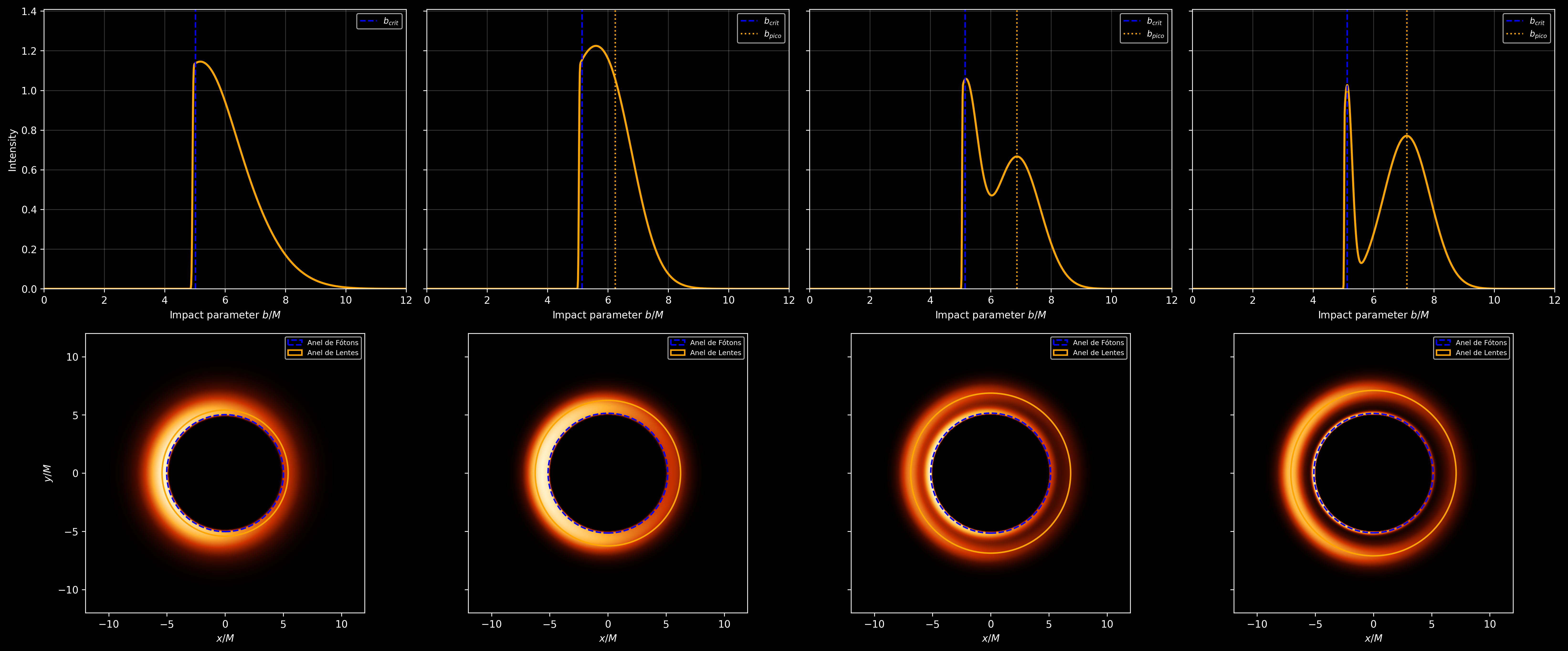} 
	\caption{Analysis of the shadow for different combinations of the parameters $a$, $n$, and $k$. Each panel demonstrates how the interaction between the parameters affects both the shadow size and the separation of the light rings, revealing the complex phenomenology of the spacetime.}
	\label{fig:comparison_multiparam}
\end{figure*}

In the previous subsections, we investigated the impact of each parameter ($a, k, n$) in isolation to understand their individual roles in the shadow morphology. In this final subsection, we analyze the combined influence of these parameters by selecting four representative cases that illustrate the phenomenology of the black-bounce spacetime. Figure \ref{fig:comparison_multiparam} presents the shadow and intensity profile for different sets of $(a, n, k)$.

Panel (a) serves as our reference case for a black-bounce with a well-defined double-ring structure, using the parameters $a=0.1 \ m$, $k=1.0$, and $n=1.0$. This case exhibits a shadow with a radius $b_{crit} = 5.0220 \ m$ and a clear separation between the photon ring and the lensing ring, whose peak is at $b_{peak} = 5.4184 \ m$.

In panel (b), we maintain $a = 0.3 \ m$, $k=2.0$, and $n=2.0$. This case exhibits a shadow with a radius $b_{crit} = 5.0220 \ m$. The dominant effect of a high $k$ is the reduction of the shadow size, which decreases to $b_{crit} = 5.1433 \ m$ and $b_{peak} = 6.2482 \ m$. The double-ring structure, governed by the parameter $a$, is preserved, but the overall scale of the image is visibly smaller.

Panel (c) explores a different regime, with a small bounce value $a = 0.6$, $k=3.0$, and $n=3.0$. Here, the dominant effect is that of the parameter $n$, which, as we have seen, tends to increase the shadow size. The result is the largest shadow among the analyzed cases, with $b_{crit} = 5.1534 \ m$. Due to the small value of $a$, the separation between the rings is almost imperceptible, and the intensity profile shows two very close peaks. This case illustrates how a large value of $n$ can "inflate" the shadow, even for a black-bounce with a modest $a$ parameter.

Finally, panel (d) shows a case of complex interaction with $a=1.0 \ m$, $n=4.0$, and $k=4.0$. The large value of $a$ ensures maximum separation between the rings, resulting in the highest value of $b_{peak} = 7.1038 \ m$. The value of $n=2.0$ contributes to an increase of the shadow compared to the base case (panel a), resulting in $b_{crit} = 5.1268 \ m$. This panel demonstrates how a large $a$ can create a large ring separation, while a high $n$ can simultaneously increase the overall size of the structure. The numerical results for all cases are presented in table \ref{tab:resultados_multiparam}.

\begin{table}[h!]
	\centering
	\caption{Numerical results for multiparametric analysis, with $m=1$.}
	\begin{tabular}{c | c | c | c | c}
		\hline
		\hline
			$a$ ($m$) & $n$ & $k$ & $b_{crit}$ ($m$) & $b_{peak}$ ($m$) \\
		\hline
		0.1 & 1.0 & 1.0 & 5.0220 & 5.4184 \\
		0.3 & 2.0 & 2.0 & 5.1433 & 6.2482 \\
		0.6 & 3.0 & 3.0 & 5.1534 & 6.8718 \\
		1.0 & 4.0 & 4.0 & 5.1268 & 7.1038 \\
		\hline
		\hline
	\end{tabular}
	\label{tab:resultados_multiparam}
\end{table}

This final analysis has demonstrated the appearance of a shadow does not depend on a single parameter, but on the non-trivial interaction between $a$, $n$, and $k$, each governing a different aspect of its morphology and size. This opens new possibilities for constraining modified gravity models through future astrophysical observations.

\section{Application of Gyoto to the shadow analysis}

\subsection{Ray-Trace Simulations with GYOTO}

To deepen our investigation and validate the results obtained with the semi-analytical model, we turn to full numerical ray-tracing simulations using the \texttt{GYOTO} code \cite{vincent2011gyoto}. \texttt{GYOTO} is an open-source software designed to integrate null and timelike geodesics a given metric. It is an ideal tool for generating realistic images of astrophysical objects in strong gravity environments.

 In simulations with \texttt{GYOTO} in order to generate more realistic images, a physical model for the light source is required. the radiation emission is generated by an accretion disk orbiting the compact object. We implemented a custom module that models a geometrically thin, optically thick accretion disk, based on the work by Page and Thorne \cite{PageThorne1974}.

The fundamental physical processes of this model are given by the disk structure. This is composed of particles (gas and plasma) that follow circular Keplerian orbits in the equatorial plane of the spacetime. As the particles in the disk lose angular momentum due to internal viscous processes, they slowly spiral toward the central object. On this journey, gravitational potential energy is converted into thermal energy, heating the disk to millions of degrees. Finally, the radiation emission takes place: this model assumes the disk to be optically thick, meaning it radiates this thermal energy efficiently. At each radius, the dissipated energy is emitted locally as blackbody radiation.

The key quantity of the model is the total (bolometric) energy flux, $F(r)$, emitted by the disk's surface at a radius $r$. This flux depends on the mass accretion rate, $\dot{M}$, and the properties of stable circular orbits in the given metric. The general formula is

\begin{equation}
	F(r) = \frac{\dot{M}}{4\pi r} \frac{-\Omega_{,r}}{(E - \Omega L)^2} \int_{r_{in}}^{r} (E - \Omega L) L_{,r'} dr',
	\label{eq:fluxo_pt}
\end{equation}
where $E$, $L$, and $\Omega$ are, respectively, the energy, angular momentum, and angular velocity of a particle in a circular orbit at radius $r$. The term $r_{in}$ represents the radius of the innermost stable circular orbit (ISCO), which acts as the disk's inner edge \cite{PageThorne1974}.

In the simulation, a simplified form of equation \eqref{eq:fluxo_pt} is used. The code first calculates the quantities $E(r)$ and $L(r)$ as seen in reference \cite{furtado2025gravitational} for our black-bounce metric. Then, it computes the flux, which is proportional to
\begin{equation}\label{pagethorne}
	F(r) \propto \frac{\dot{M}}{r^3} \frac{\sqrt{r} - \sqrt{r_{in}}}{E - \Omega L}.
\end{equation}
The term $(\sqrt{r} - \sqrt{r_{in}})$ implements the "zero torque" boundary condition at the ISCO, ensuring that the energy flux drops to zero at the inner edge of the disk.

Finally, the effective temperature of the disk at any radius, $T(r)$, is calculated from the flux using the Stefan-Boltzmann law, $F(r) = \sigma_{SB} T(r)^4$. We also use the \texttt{emission} function in simulation. It uses this temperature to calculate the radiation intensity at a specific frequency, assuming a blackbody spectrum. This physical model therefore allows us to assign a brightness and a color to each point of the accretion disk. This serves as the light source for the ray tracing given by \texttt{GYOTO}.

Considering these effects, addressed in the accretion disk and our metric, we arrive at the following results.

\subsection{Study of Black-Bounce Metric Parameters and Viewing Angle Effects Using GYOTO}

\subsubsection{Parameter $a$, $k$  and $n$}

The simulation was configured using the black-bounce metric with the free parameters $a$, $k$, and $n$. Using the function $f(r)$, we were able to determine the threshold of the event horizon, setting $f(r)=0$, so that
\begin{equation} \label{eqfr}
	f(r) = 1 - \frac{2 m r^k}{\left(r^{2n} + a^{2n}\right)^{\frac{k+1}{2n}}} \implies 1 = \frac{2 m r^k}{\left(r^{2n} + a^{2n}\right)^{\frac{k+1}{2n}}}. 
\end{equation}
This equation \eqref{eqfr} defines the horizon threshold $a_{\rm hor}$, such that by rearranging its terms, we obtain
\begin{equation}\label{eq:a2n}
    a^{2n}(r)= (2m)^{\frac{2n}{k+1}}\,r^{\frac{2nk}{k+1}}-r^{2n}.
\end{equation}

For the extremal case (double horizon), that is,  $da^{2n}/dr=0$, or, equivalently, when $f(r)=0$ and $\frac{d}{dr}f(r)=0$ simultaneously, we have
\begin{equation}\label{eq:rhor}
   r_{\rm hor}=2m\!\left(\frac{k}{k+1}\right)^{\frac{k+1}{2n}},\qquad
a_{\rm hor}=2m\!\left(\frac{k}{k+1}\right)^{\frac{k}{2n}}
\!\left(\frac{1}{k+1}\right)^{\frac{1}{2n}}. 
\end{equation}
From this, we obtain that for $a < a_{\text{hor}}$, there are two horizons (BH regime), while that for $a > a_{\text{hor}}$, there is no horizon.

In circular null orbits, the "light rings" are given by the derivative of the function $f(r)$. With this, we find the critical point $(r_\star, a_\star)$ at which the light rings disappear, it is obtained by imposing a double root condition:
\begin{equation}
    \frac{d}{dr}\!\left(\frac{\Sigma^2}{f}\right)=0,\qquad
\frac{d^2}{dr^2}\!\left(\frac{\Sigma^2}{f}\right)=0,\qquad r_\star>0,\ a_\star>0.
\end{equation}
We then observe three distinct regimes:
\begin{itemize}
\item $a < a_{\rm hor}$: \textbf{black hole} (two horizons). Typically, there is \emph{one} unstable light ring outside the horizon, and the solution projects a \emph{shadow} in the strict sense.
\item $a_{\rm hor} < a < a_\star$: \textbf{no horizon}, but \textbf{two light rings} (one unstable, one stable); imaging shows two nearly concentric thin rings and strong higher-order features.
\item $a \ge a_\star$: \textbf{no horizon and no light rings}; there is no geometric ``shadow'', and the morphology is dominated by finite throat caustics.
\end{itemize}

In addition, each light ring at $r_{\rm LR}$ determines a critical impact parameter, 
\begin{equation}\label{eq:bcrit}
b_{\rm crit}^2=\frac{\Sigma^2(r_{\rm LR})}{f(r_{\rm LR})},    
\end{equation}
which gives the observed angular radius of the thin ring (for a distant observer).

With this in mind, we were able to observe some values for the regime $a_{\text{hor}}$, $a_{\star}$, $r$, and $r_{\star}$:

\begin{center}
\begin{tabular}{c | c | c | c | c}
\toprule
$k=n$ & $a_{\rm hor}$ & $a_\star$ & $r_{\rm hor}$ & $r_\star$ \\
\midrule
\hline
1 & 1.000000 & 1.180000 & 1.000000 & 1.732051 \\
2 & 1.2408065 & 1.5963958 & 1.4757589 & 2.4984944 \\
3 & 1.3747296 & 1.8771446 & 1.6509636 & 2.8096600 \\
\bottomrule
\hline
\end{tabular}
\label{tab1}
\end{center}

Given the values in Table \ref{tab1}, we observe that for the generalized black-bounce regime, the parameter $a_{\text{hor}}$ determines the behavior of the event horizons, while $a_{\star}$ shows us the behavior of the light rings.

Applying the limits where $k=1$, $n=1$, and $a=0$ in equation \eqref{eqfr}, we return to the Schwarzschild case,
\begin{equation}
    f(r) = 1 -\frac{2m}{r}.
\end{equation}

This result confirms that the simulation represents a spherically symmetric, non-rotating black hole. The radiation source is a geometrically thin, optically thick accretion disk, modeled by equations \eqref{eqfr} and \eqref{pagethorne}, based on the Page-Thorne model. The emission flux is calculated locally, assuming a blackbody spectrum. The inner edge of the disk was set at the ISCO radius, which for the Schwarzschild spacetime is equal to $r_{\text{ISCO}} = 6m$.

\begin{figure}[h!]
	\centering
	\includegraphics[width=0.8\linewidth]{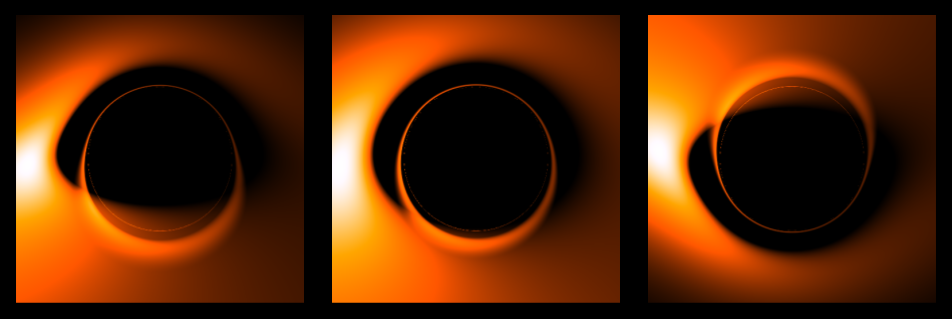}
	\caption{Images of the shadow of a black-bounce ($a=0.0 \ m, k=1.0, n=1.0$) surrounded by a thin accretion disk, generated with \texttt{GYOTO} for different observer viewing angles: (a) $\theta=20^\circ$, (b) $\theta=60^\circ$, (c) $\theta=90^\circ$. The figure illustrates the strong effect of inclination on the morphology of the accretion disk image.}
	\label{fig:screenshot001}
\end{figure}

Figure \ref{fig:screenshot001}, resulting from the simulation, displays the accretion disk image from three different observer inclination angles. Analyzing the image, we observe relativistic effects, such as the central dark region, which is the black hole shadow, whose notably circular shape is the characteristic signature of a non-rotating black hole. The intense spacetime curvature distorts the disk image, with the most evident effect being the formation of a secondary image of the back side of the disk, appearing as a bright arc above the shadow. Adjacent to its edge, an extremely thin photon ring is visible, composed of light that orbited the compact object multiple times before reaching the observer.

In the panels representing a high inclination, the asymmetry in the disk's brightness is pronounced. The side of the disk where the material approaches the observer has its flux amplified by the Doppler beaming effect, appearing brighter, while the receding side is correspondingly dimmed.

We performed the analysis for cases where $a=0.3$, $a=0.7$, and $a=1.0$, and no significant differences were observed. Therefore, to better analyze this metric, we explored other values for the bounce parameter $a$ with $k=1$ and $n=1$ fixed.

\begin{figure}[h!]
	\centering
	\includegraphics[width=0.8\linewidth]{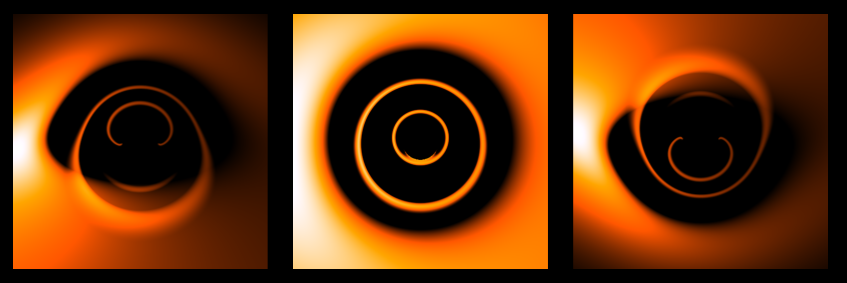}
	\caption{Images of the shadow of a black-bounce ($a=1.5 \ m, k=1.0, n=1.0$) surrounded by a thin accretion disk given by the Page-Thorne model, generated with \texttt{GYOTO} for different observer viewing angles: (a) $\theta=20^\circ$, (b) $\theta=60^\circ$, and (c) $\theta=90^\circ$. The figure illustrates the strong effect of inclination on the morphology of the accretion disk image.}
	\label{fig:a1_5}
\end{figure}

Without loss of generality, we assume larger values of $a$, starting with $a=1.5$. With this in mind, we observed that for cases where $a > 1$, the transition process from a black hole to a wormhole begins, as expected in the literature and suggested in Table \ref{tab1} and verified in Figure \ref{fig:a1_5}.

In Figure \ref{fig:a1_5}, we observe horizons, regularity, and a "throat". The horizons are given by equation \eqref{eqfr}, i.e., $r^2 - 2mr + a^2 = 0$. Thus, there are two horizons for $a < m$, a double horizon at $r = m$ when $a = m$ (extremal case), and no horizons for $a > m$ \cite{SimpsonVisser2019}\cite{Lobo2021}.

At the center $r = 0$, we have $f(0) = 1$ and a minimum area radius $\Sigma(0) = a$, meaning the geometry is free of singularities and features a throat with a minimum area of $4\pi a^2$.

The gravitational redshift is maximum at $r = a$, where $f$ reaches its minimum.

\begin{equation}
	\label{eq:fmin}
	f_{\min}=1-\frac{m}{a}
	\quad\Rightarrow\quad
	z_{\max}=\frac{1}{\sqrt{f_{\min}}}-1
	=\left(1-\frac{m}{a}\right)^{-1/2}-1.
\end{equation}

For the case of interest, $a = 1.5m$, we obtain $z_{\max} \approx 0.732$.

In black-bounce type metrics, circular null orbits satisfy \eqref{eq:fmin} and \eqref{eq:rhor}. For $a \ge a_\star$, there are no light rings. Consequently, there is no universal critical curve of the impact parameter, and therefore the solution does not project a "shadow". Any dark region at the center of synthesized images now depends on the emissive model, not on gravitational capture.

The Keplerian frequency (geodesic) for timelike circular orbits is given by

\begin{equation}\label{eq:omega}
 \Omega^2=\frac{f'}{(\Sigma^2)'}=\frac{m\,(r^2-a^2)}{r\,(r^2+a^2)^2}.   
\end{equation}

From equation \eqref{eq:omega}, it is noted that $\Omega^2 > 0$ only for $r > a$; there are no circular orbits for $r \le a$.
For a thin geodesic disk, this forces the inner edge to satisfy $r_{\rm in} \gtrsim a$ and shifts the ISCO inward.
The integrals of motion can be written as:

\begin{equation}
	\label{eq:EL}
	E=\frac{f}{\sqrt{\,f-\Omega^2\sum^2\,}},
	\qquad
	L=\frac{\sum^2\,\Omega}{\sqrt{\,f-\Omega^2\sum^2\,}},
\end{equation}

The radial stability defines the ISCO.

The renderings of a thin, inclined disk over the black-bounce metric with $a=1.5m$ exhibit:
(i) a bright crescent dominated by Doppler boosting on the approaching side;
(ii) the absence of an extremely thin critical ring (photon ring) and, in its place, relatively thick and more separated inner arcs, associated with caustics generated near the throat;
(iii) if the background source is uniform, no central shadow persists—only distortion given by the accretion disk—resulting in a regular ultracompact object without a horizon and without light rings.

\begin{figure}[h!]
	\centering
	\includegraphics[width=0.8\linewidth]{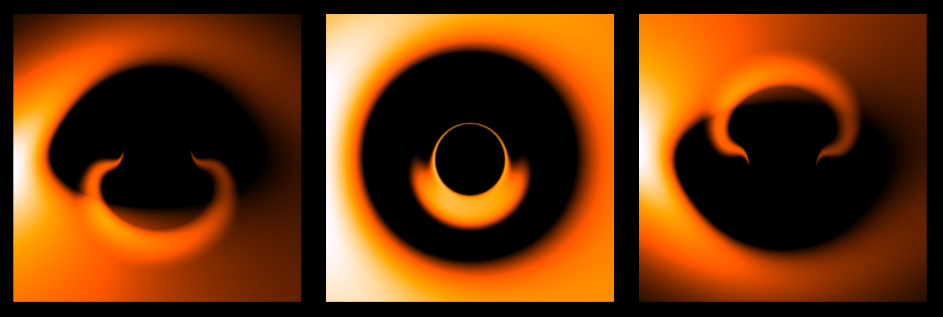}
	\caption{Images of the shadow of a black-bounce ($a=2.5 \ m, k=1.0, n=1.0$) surrounded by a thin accretion disk, generated with \texttt{GYOTO} for different observer viewing angles: (a) $\theta=20^\circ$, (b) $\theta=60^\circ$, and (c) $\theta=90^\circ$. The figure illustrates the strong effect of inclination on the morphology of the accretion disk image.}
	\label{fig:a2}
\end{figure}

Now, for the case of $a=2.0$, it falls into the regular ultracompact regime without a horizon and, since $a \ge a_\star \simeq 1.18m$, also without circular null orbits (light rings). Consequently, there is no geometric "shadow" in the strict sense. Figure \ref{fig:a2} is dominated by lensing of the throat: the Keplerian frequency given by equation \eqref{eq:omega} implies $\Omega^2 > 0$ only for $r > a$, so the thin disk has $r_{\rm in} \gtrsim a = 2m$. The thick "horseshoe"-shaped arc corresponds to the primary image of the inner edge; the additional rings/arcs are finite-order images of the same inner edge, produced by finite caustics in the vicinity of the throat.

\begin{figure}[h!]
	\centering
	\includegraphics[width=0.8\linewidth]{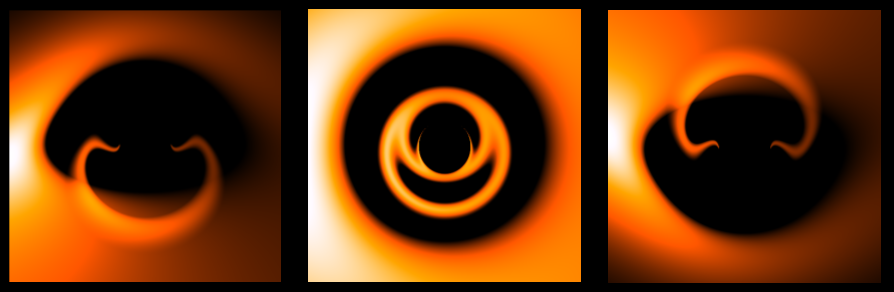}
	\caption{Images of the shadow of a black-bounce ($a=2 \ m, k=1.0, n=1.0$) surrounded by a thin accretion disk, generated with \texttt{GYOTO} for different observer viewing angles: (a) $\theta=20^\circ$, (b) $\theta=60^\circ$, and (c) $\theta=90^\circ$. The figure illustrates the strong effect of inclination on the morphology of the accretion disk image.}
	\label{fig:a2_5}
\end{figure}

For the case of $a=2.5$, we verify in Figure \ref{fig:a2_5} the absence of a horizon and light rings. Consulting Table \ref{tab1}, we find that this value corresponds to the regular ultracompact regime without a horizon and without real roots, and since $a \ge a_\star \simeq 1.18 \ m$, it also lacks circular null orbits. Therefore, there is no geometric "shadow" in the strict sense. From equation \eqref{eq:omega}, we know that $\Omega > 0$ only for $r > a$, resulting in a thin geodesic disk at $r_{\rm in} \gtrsim a = 2.5 \ m$. The images exhibit \textit{(i)} a broad central cavity due to the absence of internal emission, \textit{(ii)} a thick "horseshoe"-shaped arc corresponding to the primary image of the inner edge, and \textit{(iii)} bright cusps that mark the terminations of caustics. A thin, nearly circular ring in low-inclination views corresponds to the inner edge itself ($r_{\rm in} \sim a$), and not to a photon ring.

For the case of $a=3$ while keeping $k=n=1$, we will have a regime without a horizon and without light rings. It is in the regular ultracompact regime without a horizon because equation \eqref{eqfr} has no real roots, and since $a \ge a_\star = 1.18 \ m$, it also has no circular null orbits (no light rings). Consequently, there is no geometric "shadow" in the strict sense (absence of a universal critical impact curve).

\begin{figure}[h!]
	\centering
	\includegraphics[width=0.8\linewidth]{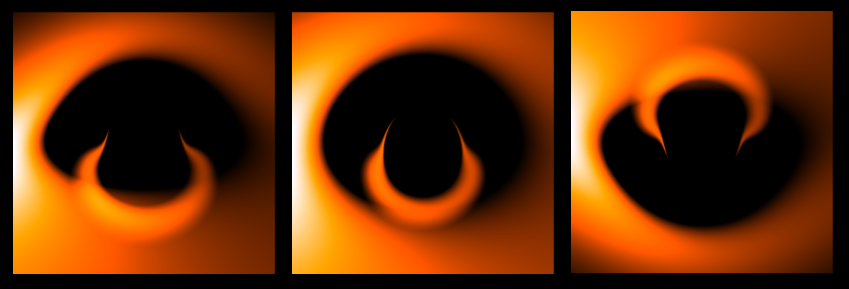}
	\caption{Images of the shadow of a black-bounce ($a=3 \ m, k=1.0, n=1.0$) surrounded by a thin accretion disk, generated with \texttt{GYOTO} for different observer viewing angles: (a) $\theta=20^\circ$, (b) $\theta=60^\circ$, and (c) $\theta=90^\circ$. The figure illustrates the strong effect of inclination on the morphology of the accretion disk image.}
	\label{fig:a3}
\end{figure}

Figure \ref{fig:a3} is dominated by gravitational lensing around the throat: the thin geodesic disk produces a broad central cavity and a thick "horseshoe"-shaped arc (primary image of the inner edge). The two bright cusps at the ends of the arc correspond to caustic terminations.

Expanding our study, we investigated the influence of the bounce parameter $a$ on the shadow morphology by fixing the exponents $k = n = 2$, values for which the critical thresholds are well established: a horizon formation threshold at $a_{\text{hor}} \approx 1.2408 \ m$ and an optical threshold at $a_* \approx 1.5964 \ m$, beyond which no circular photon orbits (\textit{light rings}) exist. We simulated two distinct scenarios, with $a$ varying from $1.5 \ m$ to $3.0 \ m$, as verified in Figure \ref{fig:a1-3kn2}.

\subsubsection*{Case $a = 1.5 \ m$ ($a_{\text{hor}} < a < a_*$)}

In this regime, the object has no event horizon but maintains two circular photon orbits: one unstable  and one stable. The optical signature of this state is unambiguous: the image exhibits two ultra-thin, nearly concentric critical rings at all inclination angles, corresponding to the \textit{light rings}. Superimposed on this fine structure, a thick, bright ring is observed, which is the primary image of the inner edge of the accretion disk. The critical impact parameters calculated for the \textit{light rings} ($b_{\text{crit}} \approx 5.79 \ m$ and $b_{\text{crit}} \approx 5.56 \ m$) are consistent with the diameters of the observed thin rings, as verified in the upper profile of Figure \ref{fig:a1-3kn2}.

\subsubsection*{Cases $a = 2.0 \ m, 2.5 \ m, 3.0 \ m$ ($a > a_*$)}

\begin{figure}[h!]
	\centering
	\includegraphics[width=0.9\linewidth]{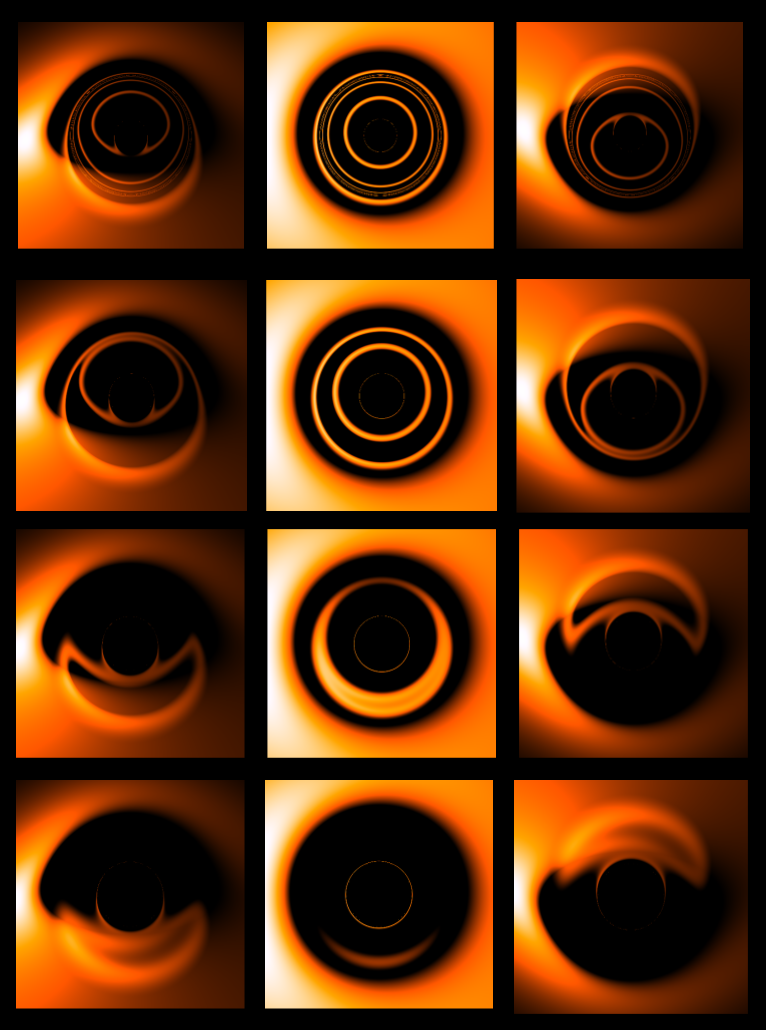}
	\caption{The figure shows images of the shadow of a black-bounce with parameters ($a=1.5 \ m, a=2.0 \ m, a=2.5 \ m, a=3.0 \ m$), respectively, with $k=n=2$ fixed, surrounded by a thin accretion disk, generated with \texttt{GYOTO} for different observer viewing angles: (a) $\theta=20^\circ$, (b) $\theta=60^\circ$, and (c) $\theta=90^\circ$. The figure illustrates the strong effect of inclination on the morphology of the accretion disk image.}
	\label{fig:a1-3kn2}
\end{figure}

For values of $a$ above the threshold $a_*$, as verified in the third row and the lower part of Figure \ref{fig:a1-3kn2}, there are no light rings. Consequently, the image is dominated by the accretion disk, characterized by:

(i) A bright, thick arc or ring, which is the primary image of the disk's inner edge.
(ii) Weaker secondary arcs, radially separated from the primary ring.

As $a$ increases, the central dark cavity expands significantly. This occurs because the inner edge radius of the geodesic disk is bounded below by $r_{\text{in}} \gtrsim 1.189a$, resulting in $r_{\text{in}} \geq 2.38 \ m$, $2.97 \ m$, and $3.57 \ m$ for $a = 2.0 \ m$, $2.5 \ m$, and $3.0 \ m$, respectively.

\begin{figure}[h!]
	\centering
	\includegraphics[width=0.9\linewidth]{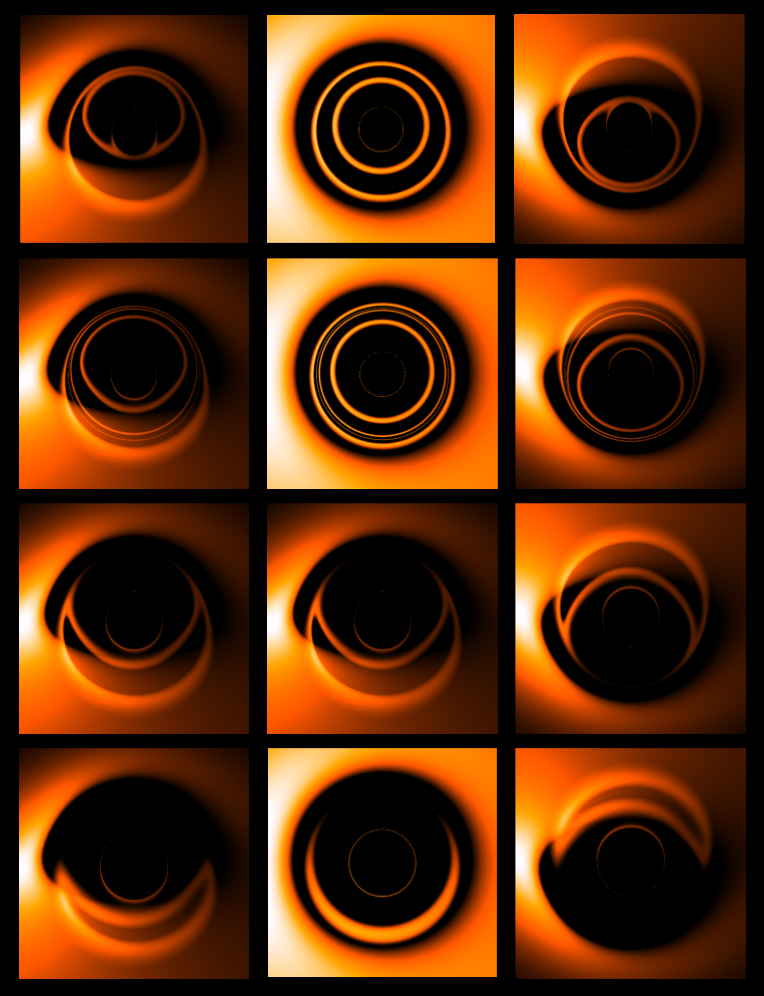}
	\caption{The figure shows images of the shadow of a black-bounce with parameters ($a=1.5 \ m$, $a=2.0 \ m$, $a=2.5 \ m$, $a=3.0 \ m$), respectively, with $k=n=2$ fixed, surrounded by a thin accretion disk, generated with \texttt{GYOTO} for different observer viewing angles: (a) $\theta=20^\circ$, (b) $\theta=60^\circ$, and (c) $\theta=90^\circ$. The figure illustrates the strong effect of inclination on the morphology of the accretion disk image.}
	\label{fig:a1-3kn3}
\end{figure}

For the case where $a = 3\ m $  and $k = n = 3 $ , the system lies above both thresholds governing strong-field optics in this family of solutions: the extremal horizon threshold  (\( a_{\text{hor}} \approx 1.3747M \)) and the disappearance threshold for \textit{light rings} (\( a_{\bullet} \approx 1.8771M \)). Consequently, the solution is regular and devoid of an event horizon, exhibiting no circular null orbits (\textit{light rings}).

The absence of \textit{light rings} implies the non-existence of a universal critical curve in the impact parameter space; therefore, there is no formation of a gravitational "shadow" in the strict sense, as seen in Figure \ref{fig:a1-3kn3}.
\clearpage

\section{Summary and Conclusions}

In this work, we analyzed the bounce parameters $a$, $k$, and $n$ and their effects on the intensity profile. We observed that as we increase the value of $a$, the function $b_{\text{crit}}$ decreases, and consequently, $b_{\text{peak}}$ increases, resulting in the formation of two well-defined peaks. For the parameter $k$, we observed a severe reduction in the shadow size, which directly affects $b_{\text{crit}}$ and $b_{\text{peak}}$ without altering the curve formation, but changing $\Delta b$. Finally, analyzing the parameter $n$, we identified the limit where $k = 2n - 1$. Thus, intrinsically, we observed that for values of $k < 4$, the function $f(r) < 1$, and therefore, no shadow is generated. Consequently, as we increase the value of $n$, the shadow size increases.

Expanding our analysis, we studied the formation of more realistic images using the \texttt{GYOTO} software, in which we implemented the accretion disk described by equations \eqref{eq:fluxo_pt} and \eqref{eqfr}. The simulation revealed two geometric thresholds that control the phenomenology: (i) the \emph{horizon} threshold $a_{\text{hor}}$, which separates solutions with and without a horizon, and (ii) the \emph{optical} threshold $a_\star$, where circular null orbits (light rings) disappear. Given these limits, the regimes are divided as follows: for $a < a_{\text{hor}}$, the solution is a black hole with an unstable light ring that defines the shadow boundary; for $a_{\text{hor}} < a < a_\star$, the spacetime is regular and horizonless but retains two light rings, manifested as two thin curves; and for $a \ge a_\star$, there are no light rings, and thus no geometric "shadow" exists.

All these regimes were addressed in this work, particularly for the cases of $a_{\text{hor}} < a < a_\star$ and $a = 2.0M, 2.5M, 3.0M > a_\star$. In the former, we observed two thin rings superimposed on the arc, generating two light rings, as shown in Figure \ref{fig:a1_5}. In the latter regime, two rings are formed, with "cusps" that can be interpreted as caustic terminations, as observed in Figure \ref{fig:a1-3kn2}. Finally, for $a \gg a_\star$, the images are entirely controlled by caustics and irregular rings without an event horizon, as seen in Figure \ref{fig:a1-3kn3}. In summary, the parameter $a$ (for fixed $k = n$) governs a sharp transition between "BH-like shadow," "two critical curves in horizonless objects," and "no-shadow regime."

\textbf{Acknowledgments}. This work was partially funded by the National Council for Scientific and Technological Development (CNPq). The work by A. Yu. P. has been partially supported by the CNPq project No. 303777/2023-0. The work by P. J. P. has been partially supported by the CNPq project No. 307628/2022-1. Ana R. M. Oliveira has been partially supported by CAPES. A.R.Q. acknowledges support by CNPq under process number 310533/2022-8. A.R.Q. work is supported by FAPESQ-PB


\section*{References}
\bibliographystyle{unsrt} 
\bibliography{bibli} 

\end{document}